\documentclass[prl, showpacs, superscriptaddress, twocolumn, floatfix]{revtex4}
\usepackage{color, graphicx, amssymb}  

\date{12/15/06}
  
\begin{document}  
\title{Magnetic response of mesoscopic superconducting rings with two   
order parameters}  
\author{Hendrik Bluhm}  
\email{hendrikb@stanford.edu}  
\affiliation{Departments of Physics and Applied Physics, Stanford  
University, Stanford, CA 94305}  
\author{Nicholas C. Koshnick}  
\affiliation{Departments of Physics and Applied Physics, Stanford  
University, Stanford, CA 94305}  
\author{Martin E. Huber}  
\affiliation{Department of Physics, University of Colorado at Denver,   
Denver, CO 80217}  
\author{Kathryn A. Moler}  
\affiliation{Departments of Physics and Applied Physics, Stanford  
University, Stanford, CA 94305}  
  
\begin{abstract}  
The magnetic response and fluxoid transitions of superconducting 
aluminum rings of various sizes, deposited under conditions likely
to generate a layered structure, show good agreement with a
two-order-parameter Ginzburg-Landau model. 
For intermediate couplings, we find metastable states that have different phase
winding numbers around the ring in each of the two order parameters. Those 
states, previously theoretically predicted, are analogous to fractional
vortices in singly connected samples with two-order-parameter 
superconductivity. Larger coupling 
locks the relative phase so that the two order parameters are only manifest
in the temperature dependence of the response. With increasing proximitization,
this signature gradually disappears.
\end{abstract}  
  

  
\maketitle  
  
Since the discovery of two-gap superconductivity in MgB$_2$
\cite{BuzeaC:RevspM}, the coexistence of two not-too-strongly coupled
superconducting order parameters (OPs) has motivated significant
theoretical work.  A striking prediction is the existence of vortices
carrying unquantized flux \cite{BabaevE:Vorwff, BabaevE:PhadpU}. 
For non-negligible Josephson coupling, those exhibit a soliton-shaped
phase difference between the two OPs
\cite{TanakaY:Solts, BabaevE:Vorwff}. 
Theoretically, such solitons may also form when current flow along a
wire causes the relative phase to unlock, leading to 
a higher current than in the phase locked state \cite{GurevichA:Phatid}.  
Most of this theoretical work is
based on a two-OP Ginzburg-Landau (GL) theory
\cite{GurevichA:Enhucf}. In the realm of superconductivity, this model
applies for both Josephson coupled bilayer systems, which are
described microscopically by a tunneling BCS Hamiltonian, and
intrinsic two-gap systems \cite{NoceC:Micebt}.  A particularly
interesting example of the latter is Sr$_2$RuO$_4$. Its 
$p_x + i p_y$ OP would imply zero energy core excitations with
non-abelian braiding statistics, which have been envisioned as a
basis for topologically protected quantum computation
\cite{StoneM:Fusrv, DasSarmaS:Prosd}.  A similar two-OP model has also been 
used to describe a liquid metallic state of hydrogen \cite{BabaevE:Assp}, 
where the electrons and protons would form two independent superfluids. 

Compared to the large number of theory papers, there is 
little experimental work on the mesoscopic structure of such two-OP systems.
In this paper, we report the observation of soliton states and other
phenomena arising from the interplay between two OPs in quasi-1D,
superconducting rings consisting of two parallel, Josephson-coupled aluminum 
layers. 
Positioning a scanning SQUID microscope \cite{BjornssonPG:Scasqi} 
over each ring individually enabled measurements of the
current, $I$, circulating the ring as a function of applied flux,
$\Phi_a$, and temperature, $T$. The ensemble of magnetic responses
includes distinct features that cannot be explained by one-OP GL, but
can be described by numerical solutions of two-OP GL.  The inferred
coupling between the two OPs depends on the ring's annulus width, $w$,
allowing us to study the crossover between intermediate and strong
coupling regimes.  
For intermediate coupling, we find metastable states with
different phase winding numbers for each OP. Those are the 1D analogue of 
unquantized vortices \cite{BabaevE:Vorwff}, and imply a soliton-shaped phase 
difference \cite{TanakaY:Solts}.
In this regime, multiple transition pathways between states give a rich
structure of hysteretic $\Phi_a$-$I$ curves.  At stronger coupling, the
system approaches the Cooper limit of complete proximitization 
\cite{DeGennesPG:Boues}: the $\Phi_a$-$I$ curves at any
temperature can be described by one-OP GL, but the existence of two
OPs is manifest in the temperature dependence of the fitted
penetration depth, $\lambda$, and the GL-coherence length, $\xi_{GL}$.

During a single, two-month-long cooldown, we characterized
the magnetic response of 40 different rings with eight annulus widths
45 nm $\le w \le$ 370 nm and radii $R$ of 0.5, 0.8, 1.2 and 2
$\mu$m. The rings were fabricated on oxidized silicon, using liftoff
lithography with PMMA resist. The 40 nm thick Al film was
deposited by e-beam evaporation at a rate of about 1 {\AA}/s and a
pressure of approximately $10^{-6}$ mBar.  During the deposition, the
rate temporarily dropped to a negligible level for about 10 min and
subsequently recovered, which most likely caused the formation of two
superconducting layers separated by an AlO$_x$ tunneling barrier.  A disk with
a radius of 2 $\mu$m had a $T_c$ of 1.6 K, representative of the bulk
film. Using $\xi_0 = 1.6~\mu$m for pure bulk Al
\cite{MeserveyR:Equpce} and $\xi_{GL}(0) \approx 70$ nm for our rings,
as derived below, we infer a mean free path of $l_e = 1.4
\xi_{GL}(0)^2/\xi_0 = 4$ nm.  The measured critical temperatures $T_c$
of the rings ranged from 1.5 to 1.9 K, depending only on $w$. 
Both the short $l_e$ and the large $T_c$ compared to
clean bulk Al indicate small grains and a (likely related) strong effect of 
oxygen impurities \cite{PettitRB:Filsae,CohenRW:Supgaf}.

Our SQUID sensor has two counterwound pickup loops and field coils 
that are used to apply a local magnetic field \cite{HuberM:Susc}. 
We position one pickup loop over a ring, record time traces
of the SQUID response while sinusoidally varying $\Phi_a$ at a few
Hz, and average hundreds to thousands of field sweeps. 
A background, measured by retracting the SQUID from the sample, is subtracted. 
The remaining signal is the flux generated by $I$, plus a residual, elliptic 
sensor background which is negligible at lower $T$ and unambiguously 
distinguishable from the ring response at higher $T$, where fluxoid 
transitions occur. Details on the technique 
will be given elsewhere \cite{KoshnickN:Flud}. 

\begin{figure}  
\includegraphics{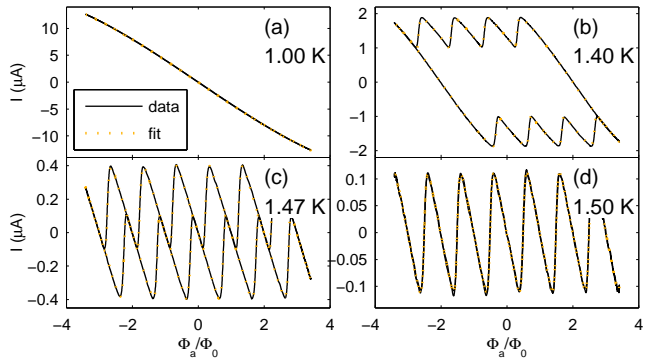}
\caption{   
\label{fig:normalPhiI} (Color online) 
$\Phi_a$-$I$ curves for a ring with $w$ = 120 nm and $R$ = 1.2 $\mu$m,
fitted to a one-order-parameter Ginzburg-Landau model. The selected
curves represent the regimes discussed in the text: (a) no
transitions, (b),(c) hysteretic and (d) thermal equilibrium.}
\end{figure}  
 
Rings with $w \le 120$ nm show no two-OP effects.
At low $T$ [Fig. \ref{fig:normalPhiI}(a)], 
there are no fluxoid transitions at the
experimental time scale and field sweep amplitude. At
higher $T$ [Fig. \ref{fig:normalPhiI}(b), (c)], we
observe  hysteretic transitions, which become non-hysteretic near $T_c$
[Fig. \ref{fig:normalPhiI}(d)].

We obtain the theoretical $\Phi_a$-$I$ curve of an individual fluxoid state, 
$n$, directly from 1D, 1-OP GL
\cite{ZhangXX:Susmsr}:
\begin{equation}\label{eq:IGL}  
I_n(\varphi) = -\frac{wd \Phi_0}{2 \pi R \mu_0 \lambda^2} (\varphi-n)  
\left(1-\frac{\xi_{GL}^2}{R^2} (\varphi-n)^2\right)  
\end{equation}    
where $\varphi = \Phi_a/\Phi_0$, $\Phi_0 = h/2e$, and $d$ is the total film 
thickness. $n$ is the  
phase winding number of the GL-OP $\psi(x) = |\psi| e^{inx/R}$, where $x$ 
is the position along the ring's circumference.  
$\varphi - n \ll R/\xi_{GL}$ gives the linear response of the  
London limit, while the cubic term arises from pair breaking.  
Because $wd \ll \lambda^2$, the self inductance can be neglected.
Although the width of some rings is several $\xi_{GL}$ at low $T$, the 
1D approximation is justified here because $w \ll R$ and $H_a \ll H_{c2}$
\cite{ZhangXX:Susmsr}.

Close to $T_c$ [Fig. \ref{fig:normalPhiI}(d)], 
transitions are fast enough to model the experimental $\Phi_a$-$I$ curves as 
a thermal average over  all possible states,  
\begin{equation}\label{eq:Ithermal}  
\langle I(\varphi)\rangle = \frac{\sum_n I_n(\varphi) e^{-E_n(\varphi)/k_BT}}  
{\sum_n e^{-E_n(\varphi)/k_BT}}  
\end{equation}  
with $E_n(\varphi) = -\Phi_0 \int_n^\varphi d\varphi' I_n(\varphi')$.
We set $\xi_{GL} = 0$ when substituting Eq. (\ref{eq:IGL}) into
Eq. (\ref{eq:Ithermal}) because the thermal rounding dominates the
cubic term
\footnote{
Fluctuation effects within a few mK of $T_c$ \cite{VonOppenF:Flupcm},
where it would be inadequate to set $\xi_{GL} = 0$, will be discussed
in Ref. \cite{KoshnickN:Flud}.
}. 
The free parameters in the fit are $\lambda^{-2}$, three background
parameters, a small offset in $\varphi$, the pickup loop--ring inductance 
$M_{coup}$, and the field coil--ring inductance.
The fitted inductances are consistent with less
accurate geometrical estimates, and are used at lower $T$. 

The hysteretic curves [Fig. \ref{fig:normalPhiI}(b), (c)] are rounded 
from averaging over a distribution of thermally activated transitions  
\cite{LangerJS:Intrtn} near a typical  $\varphi = \phi_t+n$. 
We model them by combining Eq. (\ref{eq:IGL}) 
with occupation probabilities $p_n(\varphi-n)$ obtained from integrating the 
rate equation  $dp_n/dt = -p_n/\tau_0  \exp(-E_{act}(\varphi-n)/k_B T)$.
For fitting, $\tau_0$ and the activation energy $E_{act}$ 
are absorbed into $\phi_t$. The free parameters are
$\phi_t$, $dE_{act}/d\varphi(\phi_t)$, $\xi_{GL}$,
 $\lambda^{-2}$, and three background parameters. $\varphi_t$
increases very weakly with increasing field sweep frequency,
as expected for thermally activated behavior.

For $T \ll T_c$, where no transitions are seen, Eq. (\ref{eq:IGL})
fits the data with three parameters corresponding to  $\xi_{GL}$, 
$\lambda^{-2}$, and a constant background.

\begin{figure}  
\includegraphics{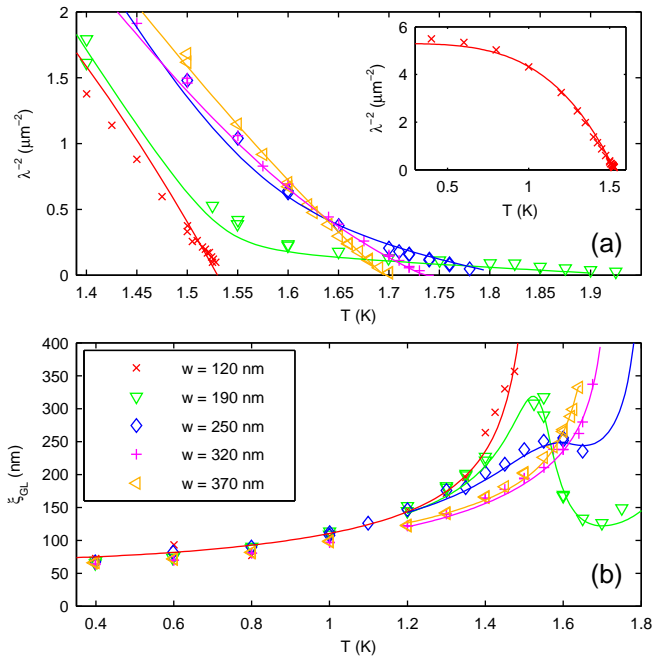}  
\caption{\label{fig:ilsXitempDep} (Color online) (a) $\lambda^{-2}$
and (b) $\xi_{GL}$ for rings representative of each annulus width
$w$. The discrete symbols are obtained from fits to $\Phi_a$-$I$
curves. Continuous curves represent fits to phenomenological
expressions ($w$ = 120 nm) or the two-order-parameter GL model ($w \ge
190$ nm), which was only fitted above 1.2 K where GL applies.}
\end{figure}

The above models result in excellent fits for measured
$\Phi_a$-$I$ curves except at $w = 190$ nm as discussed below.  The
fitted values for $\lambda^{-2}$ and $\xi_{GL}$ are shown in Fig.
\ref{fig:ilsXitempDep} for rings typical of each $w$.  All 14 measured
rings with $w \le 120$ nm show $T$ dependences similar to the 1-OP
phenomenological expressions $\lambda(T)^{-2} = \lambda(0)^{-2}
(1-t^4)$ and $\xi_{GL}(T) = \xi_{GL}(0)\sqrt{(1+t^2)/(1-t^2)}$, with
$t = T/T_c$.  For $w \ge 190$ nm, $\lambda^{-2}(T)$ has a
high-temperature tail to an enhanced $T_c$ and a peak in $\xi_{GL}$
near but below $T_c$. Both effects are most pronounced at $w=190$ nm.
The $\Phi_a$-$I$ curves remain hysteretic well into the tails, showing
that the enhanced $T_c$ is not a fluctuation effect.

\begin{figure}  
\includegraphics{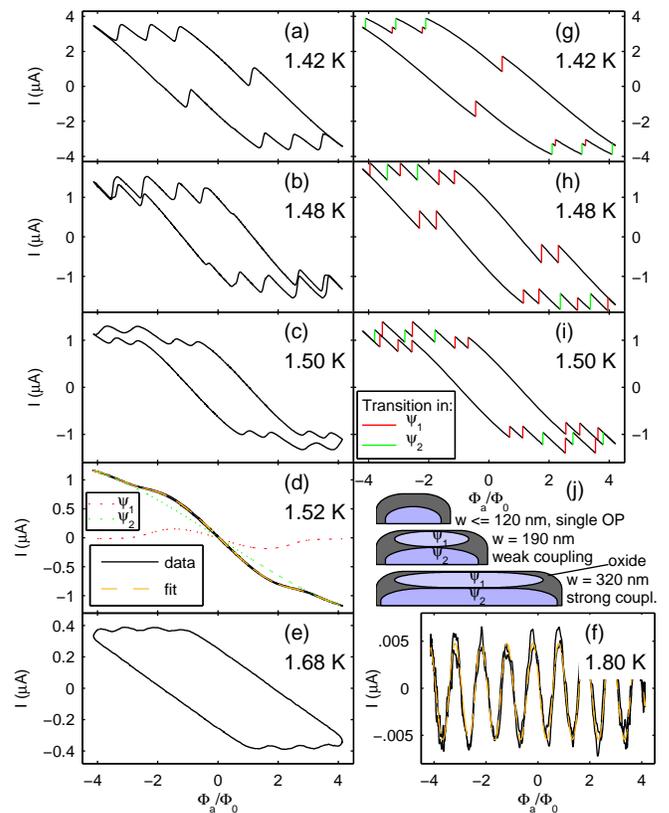}  
\caption{\label{fig:anomalousPhiI} (Color online)  
(a)-(f): $\Phi_a$-$I$ curves for a ring with $w$ = 190 nm and $R$ = 2 $\mu$m, 
where the coupling between the two order parameters is weak. In  
(a) - (c), states with two different fluxoid numbers lead to the observed  
transition pattern.  
In (d), no transition occurs because one component is stabilized by   
the other. Dotted lines show the contributions of $\psi_1$ and $\psi_2$ 
derived from the model.
(e) and (f), taken above the lower $T_c$, reflect the response of a single 
order parameter.
(g)-(i): Results  from the two-OP model corresponding to (a)-(c).  
(j): Schematic of film structure (cross section).  
}  
\end{figure}

The most striking features of the $\Phi_a$-$I$ curves for a $w=190$ nm ring 
[Fig. \ref{fig:anomalousPhiI}(a)-(f)], 
typical for all six measured rings with  $w=190$ nm and $R \ge 0.8~\mu$m
\footnote{For rings with $R = 0.5~\mu$m, $\Phi_a$ was too small to 
observe those features.},
are  transition points in the lower $T$ hysteretic 
region that are not a function of $\varphi - n$ only, and 
reentrant hysteresis. 
The latter is qualitatively related to the local maximum in $\xi_{GL}(T)$ 
[Fig. \ref{fig:ilsXitempDep}(b)], since  in 1-OP GL, a state becomes unstable
at $|\varphi - n| \ge \sqrt{R^2 /\xi_{GL}^2 + 1/2}/ \sqrt{3}$ 
\cite{VodolazovDY:Mulfja}. However,
 rather than an increase of the transitions point $\phi_t$ upon raising $T$,
 as expected for a decreasing $\xi_{GL}(T)$,  
the amplitude of those transitions is reduced until a non-hysteretic 
curve with a flattening too pronounced to be described by Eq. (\ref{eq:IGL})
[Fig. \ref{fig:anomalousPhiI}(d)] appears. 
With a further increase of $T$ [Fig. \ref{fig:anomalousPhiI}(e), (f)], 
this flattening also disappears 
and the $\Phi_a$-$I$ curves evolve similarly to Fig. \ref{fig:normalPhiI}.
This strongly suggests the existence of two OPs, one causing the small
ripples in Fig. \ref{fig:anomalousPhiI}(c), and one with a larger
$T_c$ adding the large background response and causing the tail in
$\lambda^{-2}(T)$. The additional phase winding number from
a second, coupled OP also explains the irregular transitions
in [Fig. \ref{fig:anomalousPhiI}(a)-(c)].
Although also breaking strict flux periodicity, finite line 
width corrections to 1D, 1-OP GL \cite{ZhangXX:Susmsr}  
would be much smaller and more regular. 
A vortex pinned in the annulus could in principle lead to similar 
$\Phi_a$-$I$ curves, however $w$ is too small compared to 
$\xi_{GL}$ to accommodate 
its core,and it cannot explain the $T$ dependence of $\lambda^{-2}$ 
and $\xi_{GL}$.

A two-OP GL-model consisting of two standard 1D GL free energy functionals   
and a coupling term indeed reproduces the peculiar features of  
Fig. \ref{fig:ilsXitempDep} and \ref{fig:anomalousPhiI}:  
\begin{eqnarray}\label{eq:twoCompGl}  
F[\psi_1, \psi_2, \varphi] &=& F_1[\psi_1, \varphi] + F_2[\psi_2, \varphi]   
\nonumber\\  
&+& \frac {\gamma w} 2 \int_0^L dx|\psi_1 - \psi_2|^2 \quad\textrm{with}\\  
F_i[\psi_i, \varphi] &=& w d_i \int_0^L dx \left\{\frac{\hbar^2}{2m}  
\left|\left(-i\nabla + \frac \varphi R\right)\psi_i\right|^2 \right.
\nonumber\\  
&+& \left.\frac{\alpha_i}{2} |\psi_i|^2 + \frac{\beta_i}{4} |\psi_i|^4\right\}
\end{eqnarray}
We define $\psi_1$ to have the lower $T_c$.
If both components have the same $n$, one can make the usual ansatz 
$\psi_i(x) = |\psi_i|e^{inx/R}$. 
Minimizing (\ref{eq:twoCompGl})  with respect to $|\psi_1|$ and $|\psi_2|$
results in excellent fits to $\Phi_a$-$I$ curves as in Fig. 
\ref{fig:anomalousPhiI}(d). At small $\varphi$, both OPs
contribute significantly, but as $\varphi$ is increased, pair breaking
strongly reduces $|\psi_1|$, whose $\xi_{GL}$ diverges near $T_{c,1}$.
In the absence of the more
stable $\psi_2$, $\psi_1$ would undergo a fluxoid transition 
much before $|\psi_1|$ could be suppressed that much.

We extracted effective values of $\lambda^{-2}(T)$ and
$\xi_{GL}(T)$ from such modeled $\Phi_a$-$I$ curves (and also from
fits to datasets similar to Fig. \ref{fig:anomalousPhiI}(d)) using a small
$\varphi$ expansion analogous to Eq. (\ref{eq:IGL}). 
Assuming a linear $T$ dependence of $\alpha_1$ and $\alpha_2$, 
this procedure can reproduce the observed form of $\lambda^{-2}(T)$ and 
$\xi_{GL}(T)$ as demonstrated by the fits in Fig. \ref{fig:ilsXitempDep}.  
The knee in $\lambda^{-2}(T)$ corresponds to the lower   
$T_{c,1}$, above which the amplitude of $\psi_1$ becomes very small.   
From $w$ = 190 nm to $w$ = 370 nm, the coupling strength $\gamma$ 
increases by a factor 30-50, by far the largest line width dependence 
of all fit parameters
\footnote{See auxiliary document at \\
\texttt{http://www.stanford.edu/group/moler/publications.html} for a 
discussion of the fit procedures and parameters.}.
The resulting stronger proximitization smears out the two-OP features.
It appears that $\psi_1$ not only has a lower $T_c$, but also a 
smaller $l_e$ than $\psi_2$.
Fit parameters obtained from $\Phi_a$-$I$ curves and the curves in 
Fig. \ref{fig:ilsXitempDep} are consistent within about a factor 2.
This discrepancy may be due to fluctuations 
of $\psi_1$ at large $\varphi$, or a nontrivial phase difference 
\cite{GurevichA:Phatid}, which are not considered in our 
model for $\Phi_a$-$I$ curves.
 
To model the hysteretic $\Phi_a$-$I$ curves 
[Fig. \ref{fig:anomalousPhiI}(a)-(c)],   
we calculate the activation energies $E_{act}(\varphi-n)$ for transitions  
in either $\psi_1$ or $\psi_2$ 
by numerically computing saddle point solutions of the GL - equations  
obtained from variation of (\ref{eq:twoCompGl}), assuming that the   
relevant saddle  
points evolve continuously from those of the uncoupled system upon  
increasing $\gamma$.  We then derive the complete $\Phi_a$-$I$ curve  
assuming that a transition in $\psi_i$ occurs whenever   
$E_{act}(\varphi-n) <  \kappa_i k_B T$, where the $\kappa_{1,2}$ are 
treated as phenomenological parameters of order unity.
Of several simulation runs with various parameters similar to those obtained
from the fits, the one shown in Fig. \ref{fig:anomalousPhiI}(g)-(i)
gave the best similarity with the data [Fig. \ref{fig:anomalousPhiI}(a)-(c)].
While the uncertainty of the fit parameters \cite{endnote25} and 
simplicity of the model forbid 
a more quantitative comparison, the simulations show that 
metastable states with $n_1 \neq n_2$ are key to understanding
the observed $\Phi_a$-$I$ curves. Due to the coupling between
the OPs, the variation of the relative phase is soliton-like
\cite{TanakaY:Solts}, similar to a Josephson vortex, 
however with the phase gradient along the junction being mostly due to
the kinetic rather than the magnetic inductance.  
This soliton corresponds to the domain walls originating from unquantized 
vortices in bulk samples \cite{BabaevE:Vorwff}. 
Intuitively, it is formed upon increasing $\varphi$ when the larger $\xi_{GL}$
of $\psi_1$ causes $\psi_1$ to become unstable at a smaller $\varphi$ than 
$\psi_2$, and the coupling is weak enough for $\psi_2$ to stay in the
same state. 
However, the soliton energy 
makes it less stable than states with $n_1 = n_2$.  This leads to the 
observed step-like transition sequence.

Apart from the general agreement with the simulations and the irregular 
transitions, the most
direct experimental evidence for states with  $n_1 \neq n_2$ are branches 
of $\Phi_a$-$I$ curves
that are shifted relative to each other by less than one $\Phi_0$ horizontally,
such as around $\Phi_a/\Phi_0 \approx \pm 3$ in Fig. 
\ref{fig:anomalousPhiI}(b).
Individual, unaveraged field sweeps show that this is not an effect of 
averaging over different transition pathways, contrary to the features at
$\Phi_a/\Phi_0 \approx \pm 0.3$.

The emergence of two OPs can be explained by the
temporary drop of the deposition rate during the metalization. At the
lower rate, more oxygen was co-deposited to form a
tunneling barrier [see Fig.  \ref{fig:anomalousPhiI}(j)].  Different
oxygen concentrations and/or grain sizes led to different
values of $T_c$ and $\xi_{GL}$ in the two superconducting layers.
It is known that PMMA outgases significantly and that 
thinner lines are affected more \cite{DubosP:Thetrn}.
This likely caused the line width dependence of
the coupling and the complete oxidization of one of the
superconducting layers for $w \le 120$ nm, where we found no evidence
for two OPs.  
The critical Josephson current densities estimated from the inferred values
of $\gamma$ support this picture \cite{endnote25}.

This fabrication result was unintentional but fortuitous. 
While the parameters should be tunable a priori with 
controlled exposure to oxygen gas, it would be difficult to obtain
reproducible results from outgasing resist.
Nevertheless, the data and analysis draw a clear picture of a 
two-OP superconductor with GL parameters that 
depend consistently on $w$. This dependence and the occurrence
of two different $T_c$'s allowed the study of a wide range of parameters.
The insight thus gained may be used to design similar experiments on intrinsic 
two-component superconductors. The creation and
detection of $h/4e$ vortices in Sr$_2$RuO$_4$ would be of particular 
interest \cite{StoneM:Fusrv, DasSarmaS:Prosd}. 
Since their energy is logarithmically or linearly 
divergent in the sample size \cite{BabaevE:Vorwff, BabaevE:PhadpU}, 
they might only be accessible as metastable states in mesoscopic samples,
similar to the soliton states discussed here.

In conclusion, we have explored effects emerging from two coupled order 
parameters in mesoscopic
superconducting rings.  The most interesting ones
are an anomalous temperature dependence of the average superfluid
density $\lambda^{-2}$ and effective coherence length $\xi_{GL}$, and a 
qualitative modification of the behavior of phase slips related to previously 
predicted metastable states with two different phase winding numbers 
\cite{GurevichA:Phatid, BabaevE:Vorwff} and a soliton-shaped phase difference
\cite{TanakaY:Solts}.  
\acknowledgments{This work was supported by NSF Grants No.
DMR-0507931, DMR-0216470, ECS-0210877 and PHY-0425897 and by the 
Packard Foundation. Work was performed in part at the Stanford 
Nanofabrication Facility, which is supported by NSF Grant No. ECS-9731293, 
its lab members, and industrial affiliates. We would like to thank 
Per Delsing, Egor Babaev and Mac Beasley for useful discussions.}
  
\bibliography{two_comp_bibdata,pc_bibdata}  
  
\end{document}